\documentclass{jetpl}
\twocolumn
\title{Nambu sum rule in the NJL models: from superfluidity  to top quark condensation
}

\rtitle{Nambu sum rule in the NJL models: from superfluidity to top quark condensation
}

\sodtitle{Nambu sum rule in  the NJL models: from superfluidity to top quark condensation
}

\author{G.E. Volovik $^{*+}$
\/\thanks{e-mail: volovik@boojum.hut.fi}
and M.A. Zubkov $^{**}$\/\thanks{e-mail: zubkov@itep.ru}
}

\rauthor{G.E.Volovik, M.A.Zubkov}

\sodauthor{Volovik, Zubkov}

\address{
$^{*}$ O.V. Lounasmaa Laboratory, Aalto University, School of Science and
Technology, P.O. Box 15100, FI-00076 AALTO, Finland
\\
$^+$ Landau Institute for Theoretical Physics RAS, Kosygina 2,
119334 Moscow, Russia
\\
$^{**}$ ITEP, B.Cheremushkinskaya 25, Moscow, 117259, Russia
  }

\dates{\today}{*}


\abstract{
It may appear that the recently found resonance at 125 GeV is not the only
Higgs boson.
 We point out the possibility that the Higgs bosons appear in models of
top-quark condensation, where the masses of the bosonic excitations are
related to the top quark mass by the sum rule similar to the Nambu sum rule
of the NJL models \cite{Nambu}. This rule was originally considered by Nambu
for superfluid $^3$He-B and for the BCS model of superconductivity.  It
relates the two masses of bosonic excitations existing in each channel of
Cooper pairing to the fermion mass. An example of the Nambu partners is
provided by the amplitude and the phase modes in the BCS model describing
Cooper pairing in the s-wave channel.  This sum rule suggests the existence
of the Nambu partners for the 125 GeV  Higgs boson. Their masses can be predicted by the Nambu sum rule
under certain circumstances. For example, if there are only two states in the given channel, the mass of the Nambu partner is $\sim$ 325 GeV.  They together satisfy the Nambu sum rule  $M_1^2 +
M_2^2 = 4 M_t^2$, where $M_t \sim $174 GeV is the mass of the top quark. If there are two doubly degenerated states, then the second mass is $\sim$ 210 GeV.  In this case the Nambu sum rule is $2 M_1^2 + 2 M_2^2 = 4 M_t^2$.
In addition, the properties of the Higgs modes in superfluid $^3$He-A, where
the symmetry breaking is similar to that of the Standard Model of particle
physics, suggest the existence of two electrically charged Higgs particles
with masses around 245 GeV, which together also obey the Nambu sum rule
$M_+^2 + M_-^2 = 4 M_t^2$.
}

\begin{document}

\maketitle

\section{Introduction}

In 1985 Nambu noticed the relation between the energy gaps of bosonic and
fermionic excitations in a certain class of  the effective NJL - like
models (i.e. the models with the 4 - fermion interaction)  \cite{Nambu}.
This class includes superfluid $^3$He-B and  $s$-wave  superconductors.
The collective bosonic modes emerging in the fermionic system (Goldstone and
Higgs bosons) can be distributed into the pairs of Nambu partners. For each
pair one has the relation,
 $M_{1}^2 + M_{2}^2 = 4 M^2_f$, where $M_1$ and $M_2$ are gaps in the
bosonic spectrum, and
 $M_f$ is the gap in the femionic spectrum (in relativistic systems they
correspond
 to the masses of particles).
 The similar relation was also discussed in the Nambu - Jona - Lasinio (NJL)
approximation \cite{NJL} of QCD, where  it relates the $\sigma$ - meson mass
and the constituent quark mass $M_{\sigma} \approx 2 M_{quark}$.

Here we suggest that Higgs bosons in the Standard Model are composite
objects, and
they obey the same relation which we call the Nambu sum rule
 \begin{equation}
 \sum M_{H, i}^2  \approx 4 M^2_f\,.
  \label{NSR_}
\end{equation}
Here $M_{H, i}$ are the masses of composite  Higgs bosons  within the given
channel,
and $M_f$  is the mass of the heaviest  fermion, which contributes to their
formation. We assume that this is the top quark.

We suggest the hypothesis that  Eq. (\ref{NSR_}) holds in the theories that admit the NJL approximation if there is the fermion
whose mass $M_f$ dominates the fermion spectrum. We
apply this sum rule for the estimation of the masses of extra Higgs bosons,
since
the analogy with the superconductivity and superfluidity  prompts that the
Higgs
boson may be composite\footnote{See \cite{Englert,Higgs,Kibble} for the
foundation
of the Higgs mechanism in quantum field theory.}.

It is worth mentioning that the particles within the masses larger than
$130$ GeV are not excluded by present experiments if they have the cross -
sections smaller than that of the standard Higgs boson of the Standard Model
\cite{CMSHiggs, ATLASHiggs}. For example, on Figure 4 of \cite{CMSexc} the
solid black curve
separates the region, where the scalar particles are excluded (above the
curve) from the region, where they are not excluded. In particular, the
particle with
mass around $200$ GeV and with the cross section about $1/3$ of the Standard
Model cross section is not excluded by these data\footnote{The similar
exclusion
curve
was announced by ATLAS (plenary talk \cite{ATLASexc} at ICHEP 2012, slide
34).}.

 The Nambu sum rule  Eq. (\ref{NSR_}) gives an important constraint on the
bosonic spectrum. For example, if there are only two states in the channel
that contains the discovered $125$ GeV Higgs boson, then the partner of this
boson should have the mass around $325$ GeV.
 Surprisingly, in 2011 the
CDF collaboration \cite{CDF} has announced the preliminary results on the
excess of events in $Z Z \rightarrow l l \bar{l} \bar{l}$ channel at the
invariant mass $\approx 325$ GeV.  CMS collaboration also reported a small
excess in this region \cite{CMS}.
In \cite{325,320} it was argued that this may point out
to the possible existence of  a new scalar particle with mass $M_{H2}\approx
325$ GeV.
If there are two (doubly degenerated) Higgs bosons in the channel that contains the $125$ GeV Higgs, then the partner of the $125$ GeV boson should have mass around   $210$ GeV. (This possibility is realized in the model of Section 5 of the present paper.)
In the channel with two states
of equal masses the $245$ GeV Higgs
bosons should appear in analogy with $^3$He-A considered in Section 3.
 A certain excess of events in this region has been
observed by ATLAS
in 2011 (see, for example, \cite{ATLAS}).

We review several models, where the Nambu sum rule holds.
The paper is organized as follows. In Section 2  we consider the appearance
of the Nambu Sum rule in $^3$He-B mentioned originally by Nambu. In Section
3 we consider the 3D A-phase of the superfluid $^3$He. In this case the
fermions
are gapless. However, the Nambu sum rule Eq. (\ref{NSR_}) works if in its
r.h.s. the average of the angle dependent energy gap is substituted.  In
Section 4 we review bosonic
excitations and Nambu sum rules in the 2D thin films of He-3. There are two
main phases, where the
Nambu sum rule works within the effective four - fermion model similar to
that of $^3$He-B.
In section 5 we consider how the sum rule Eq. (\ref{NSR_}) appears in its
nontrivial form in the relativistic NJL model. Namely, we deal with the
particular case considered in \cite{VolovikZubkov2012} of the model of the
top -
quark condensation suggested in \cite{Miransky}. (This model is the direct
generalization of the original  model of \cite{topcolor1} to the case, when
all $6$ quarks are included.) In Section 6 we review results on the bosonic
excitations in dense quark matter. We consider diquarks in Hadronic phase
and  the color superconductor in the so - called Color - Flavor  Locking
(CFL) phase.
In Section 7 we compare Veltman relation for the vanishing of quadratic
divergences to the scalar boson mass with the Nambu sum rule. In Section 8
we end with the conclusions.

\section{Nambu sum rules in $^3$He-B}

In the B - phase of $^3$He the condensate is formed in the spin-triplet
$p$-wave state,
which is characterized by the quantum numbers of spin, orbital momentum and
total angular momentum correspondingly
$S=1$, $L=1$,
$J=0$ \cite{VollhardtWolfle1990}. This corresponds to the symmetry breaking
scheme $G \rightarrow  H$  with
the symmetry of physical laws $G=SO_S(3) \times SO_L(3) \times U(1)$ and the
symmetry of the degenerate vacuum states $H=SO_J(3)$.
The order parameter is $3\times 3$ complex matrix
\begin{equation}
A_{\alpha i} = \Delta \delta_{\alpha i} + u_{\alpha i} +iv_{\alpha i} \,.
\label{B-phase}
\end{equation}
Here the first term corresponds to the equilibrium state, with $\Delta$
being the gap in the fermionic spectrum. The other two terms
are the deviations from the equilibrium. They represent 18 collective
bosonic modes, which are
classified by the total angular momentum quantum number $J = 0,1,2$.
At each value of $J=0,1,2$ the  modes $u$ and $v$ are orthogonal to each
other and correspond to different values of the bosonic energy gaps.
Four modes are gapless. They represent Goldstone bosons, which result from
the symmetry breaking. The other  14 modes are Higgs bosons with non-zero
gaps. \footnote{Higgs bosons in other condensed matter systems
have been discussed in recent papers  \cite{Gazit2012,Barlas2012,KunChen2013} and in references therein.}

The energy gaps of bosons in $^3$He-B are given by:
\begin{equation}
E_{u,v}^{(J)} = \sqrt{ 2 \Delta^2(1\pm  \eta^{(J)})}  \,,
\label{SymmetryConsideration}
\end{equation}
where parameters $ \eta^{(J)}$ are determined by the symmetry of the system,
 $\eta^{J=0} = \eta^{J=1} = 1$,
and $\eta^{J=2} = \frac{1}{5}$
 \cite{VolovikZubkov2012}.
Equation \eqref{SymmetryConsideration} proves the sum rule for $^3$He-B
found by Nambu for $^3$He-B:
\begin{equation}
[E_{u}^{(J)}]^2 + [E_{v}^{(J)}]^2  = 4 \Delta^2 \,.
\label{NSRThe}
\end{equation}

For $J=0$ there is one pair of the Nambu partners (the gapless Goldstone
sound mode and the Higgs mode, which is called the pair-breaking mode):
\begin{equation}
E^{(0)}_1 = 0,\quad E^{(0)}_2 =  2 \Delta.
\end{equation}

For $J=1$ there are three  pairs of Nambu partners (three gapless Goldstone
modes --  spin waves, and three Higgs  pair-breaking modes):
\begin{equation}
E^{(1)}_1 = 0,\quad E^{(1)}_2 =  2 \Delta.
\end{equation}

For $J=2$ there exist five pairs of Higgs partners -- five so-called real
squashing modes
with the energy gap $E^{(2)}_1$, and, correspondingly,  five
imaginary squashing modes
   with the energy gap $E^{(2)}_2$:
  \begin{equation}
E^{(2)}_1 = \sqrt{2/5}\, (2\Delta),\quad E^{(2)}_2 =  \sqrt{3/5}\,
(2\Delta)\,.
\label{He3B2}
\end{equation}
Zeeman splitting of imaginary squashing mode in magnetic field has been
observed in \cite{Lee1988},
for the latest experiments  see  \cite{Collett2012}.

\section{Nambu sum rules in bulk $^3$He-A}

In the A-phase of $^3$He the condensate is formed in the state with
$S_z=0$ and $L_z=1$ \cite{VollhardtWolfle1990}.  In the orbital sector the
symmetry breaking in $^3$He-A is similar to that of the electroweak theory:
$U(1)\otimes SO_L(3)\rightarrow U_Q(1)$, where the quantum number $Q$  plays
the role of the electric charge (see e.g. Ref.
\cite{VolovikVachaspati1996}), while in the spin sector one has
$SO_S(3)\rightarrow SO_S(2)$.
The order parameter matrix has the form
\begin{equation}
A_{\alpha i} = \Delta_0 {\hat z}_{\alpha} ({\hat x}_i +i{\hat y}_i)+
u_{\alpha i} +iv_{\alpha i} \,.
\label{A-phase}
\end{equation}
The A-phase is anisotropic. The special direction  in the orbital  space
appears that is identified with the direction of the spontaneous orbital
angular momentum of Cooper pairs, which is here chosen along the axis $z$.
In this phase fermions are gapless: the gap in the fermionic spectrum
depends on the angle between
the momentum ${\bf k}$ and  the
anisotropy axis, $\Delta(\theta) = \Delta_0 \sin \theta$,  and nullifies at
$\sin \theta=0$.
The spectrum of the  collective modes has been considered in
\cite{BrusovPopov1980},
see also \cite{VolovikKhazan1982} for extra Goldstone modes related to
hidden symmetry of the A-phase.
The energy spectrum of the Higgs modes has, in general, the imaginary part
due to the
radiation of gapless fermions. However, if the radiation processes are
neglected, one obtains  that there are the Nambu partners that satisfy a
version of the Nambu sum rule,  written in the form
\begin{equation}
E^2_1 + E^2_2 = 4 \bar{\Delta}^2
\label{ClappingModes}
\end{equation}
The role of the square of the fermion mass is played by the
angle average of the square of the anisotropic gap in the fermionic spectrum:
\begin{equation}
\bar{\Delta}^2 \equiv  \left<\Delta^2(\theta)\right> =\frac{2}{3}\Delta_0^2
\,.
\label{FermionMass}
\end{equation}

One  (triply degenerated) pair of bosons (the phase and amplitude collective
modes in Nambu terminology) is formed by the ``electrically neutral''
($Q=0$) massless Goldstone mode and the  ``Higgs boson''  with $Q=0$:
\begin{equation}
E^{(Q=0)}_1 = 0,\quad E^{(Q=0)}_2 =  2 \bar{\Delta}   = \sqrt{8/3}
{\Delta_0}. \label{A1}
\end{equation}
The other  (triply degenerated) pair represents the analog of the charged
Higgs bosons  in $^3$He-A with $Q=\pm 2$.
These are the so-called clapping modes whose energies are
\begin{equation}
E^{(Q=2)}_1 = E^{(Q=-2)}_2 =  \sqrt{2} \bar{\Delta} = \sqrt{4/3} {\Delta_0}.
\label{Q2}
\end{equation}

\section{Superfluid  phases in 2+1 films}

The same relations \eqref{A1} and \eqref{Q2} take place for the bosonic
collective modes in the quasi two-dimensional superfluid $^3$He films. There
are two possible phases in thin films,  the A-phase and the planar phase.
Both phases have isotropic gap $\Delta$ in the 2D case, as distinct from the
3D
case where such phases are anisotropic with zeroes in the gap, and both have
similar
spectrum of 12 collective modes: 3  Goldstone bosons and 9 Higgs modes.
The energy gaps of bosons are given by equation
\eqref{SymmetryConsideration},
where instead  of $J$ there is the corresponding quantum number. This proves
that
the collective modes obey the Nambu sum rule.
The parameters $\eta$ are determined by the symmetry of the system, but
in both cases they get three possible values
  $\eta=1$, $\eta= -1$,
and $\eta = 0$.

Let us enumerate the modes in the thin film of A-phase,  where the symmetry
breaking is
$SO(2) \otimes SO(3) \otimes U(1) \rightarrow U(1)_Q\otimes SO(2)$, and the
bosonic modes are classified in terms of the
$U(1)$ charge $Q$, which is similar to the electric charge in Standard
Model.
These modes  form two pairs of Nambu partners (triply
degenerated), with $Q=0$ and $|Q|=2$:
\begin{eqnarray}
E_1^{(Q=0)} =0~~,~~ E_2^{(Q=0)} = 2 \Delta\,,
\label{2DA1}
\\
E^{(Q=+2)}=
\sqrt{2}\Delta~~,~ E^{(Q=-2)}= \sqrt{2}\Delta  \,.
\label{2DA2}
\end{eqnarray}
This spectrum was originally obtained in Ref. \cite{BrusovPopov1981}.

Note that since masses of $Q=+2$ and $Q=-2$ modes are equal,
the Nambu sum rule necessarily leads
to the definite value of the masses of the ``charged'' Higgs bosons.
Because of the common symmetry breaking scheme in the electroweak theory and
in $^3$He-A we consider the listed above energy gaps as an indication of the
existence of the Higgs boson with mass
\begin{equation}
M_H=\sqrt{2} M_t  \,.
\label{2D}
\end{equation}
This mass is about 245 GeV.

\section{Nambu sum rules in the relativistic models of top quark
condensation}

In this section we consider the Nambu sum rule in the context of the
extended NJL model of top - quark condensation.
The simplest models of this kind were considered in a number of papers (see,
for example,
\cite{topcolor1,2HMC,2HMMiransky}).  Here we consider the particular case of
the model suggested by Miransky and coauthors in \cite{Miransky}. It
involves 6 quarks and  has the action of the form
\begin{eqnarray}
S & = & \int d^4x \Bigl(\bar{\chi}[ i \nabla \gamma ]\chi \nonumber\\&& +  \frac{8\pi^2}{N_C \Lambda^2} (\bar{\chi}_{\alpha A,L} \chi^{\beta B}_R)(\bar{\chi}_{\bar{{\beta}}
\bar{B}, R} {\chi}^{\bar{\alpha} {A}}_{L}) L_{\bar{\alpha}}^{\alpha}
R_{\beta}^{\bar{\beta}} I_B^{\bar{B}}\Bigr) \label{Stopcolor_}
\end{eqnarray}

Here $\chi_{\alpha A}^T = (u,d); (c,s); (t,b)$ is the set of the doublets
with the generation index $\alpha$, $\Lambda$ is the dimensional parameter, $N_C=3$.
Hermitian matrices  $L,R,I$  contain dimensionless coupling constants.
  It is implied that all eigenvalues of matrices $L,R,I$ are close to each
other. This means that the unknown microscopic theory should have the
approximate symmetry, which provides that these values are equal. Small
corrections to this equality gives the eigenvalues of $L,R,I$ that only
slightly deviate from each other. (After the suitable rescaling $\Lambda$ plays the role of the cutoff,
while the eigenvalues of $L,R,I$ are all close to $1$.)
 The possible origin of this pattern was
discussed in \cite{Z2013}, where it is suggested that the given NJL model
originates from the gauge theory of Lorentz group coupled in an equal way to
all existing fermions.
The basis of observed quarks corresponds to the diagonal form of $L,R,I$. We
denote
$L = {\rm diag}(1+L_{ud},1+L_{cs},1+L_{tb})$, $R = {\rm
diag}(1+R_{ud},1+R_{cs},1+R_{tb})$,
 $ I = {\rm diag}(1+I_{up},1+I_{down})$, and
\begin{eqnarray}
 &&y_u  = L_{ud} +R_{ud}+ I_{up}, \quad y_d = L_{ud}+ R_{ud}
+I_{down},\nonumber\\&&
  y_c  = L_{cs}+ R_{cs} + I_{up},\quad y_s = L_{cs} +R_{cs}+
I_{down},\nonumber\\&&
  y_t = L_{tb}+ R_{tb} + I_{up}, \quad y_b = L_{tb}+ R_{tb} +I_{down},
\nonumber\\ &&
y_{ud}  = L_{ud} +R_{ud}+ I_{down}, \quad y_{du} = L_{ud} +R_{ud}
+I_{up},\nonumber\\ &&
 y_{uc}  = L_{ud} +R_{cs}+ I_{up}, \quad y_{cu} = L_{cs} +R_{ud} +I_{up},
\nonumber\\&&
  y_{us} = L_{ud}+ R_{cs}+ I_{down}, \quad y_{su} = L_{cs} +R_{ud}+ I_{up},
...\nonumber\\
&& ...\label{parameters}
\end{eqnarray}

These coupling constants satisfy the relation $y_{q_1q_2}+y_{q_1q_2} =
y_{q_1}+y_{q_2}$.
As it was mentioned above, it is implied that $|y_{q}|, |y_{q_1 q_2}| << 1$.
Bosonic spectrum of this model was calculated in one - loop approximation in
\cite{VolovikZubkov2012}. It is implied that in vacuum the composite scalar
fields $h_q = \bar{q}q$ are condensed for all quarks $q=u,d,c,s,t,b$. The
induced quark masses $M_q$ are related to the coupling constants $y_q$, $\Lambda$ as
$\frac{M_q^2}{\Lambda^2}{\rm log} \frac{\Lambda^2}{M_q^2} = y_q$.

 As a
result we have two excitations in each $q\bar{q}$ channel:
\begin{equation}
M^P_{q\bar{q}} = 0; \quad M^S_{q\bar{q}} = 2 M_q
\end{equation}
 and four excitations (i.e. two doubly degenerated excitations) in each
$q_1\bar{q}_2$ channel. We denote the masses $M^{\pm}_{q_1q_2},
M^{\pm}_{q_2q_1}$ for  $ q_1, q_2 = u,d,c,s,t,b$. They are given by
\begin{eqnarray}
M_{q_1 {q}_2}^2 &= &M_{q_1}^2 + M_{q_2}^2  \nonumber\\&&\pm
\sqrt{(M_{q_2}^2 - M_{q_1}^2)^2\zeta_{q_1q_2}^2+ 4 M_{q_1}^2M_{q_2}^2}
\end{eqnarray}
with
 \begin{equation}
 \zeta_{q_1 q_2} = \frac{2 y_{q_1q_2}- y_{q_2} - y_{q_1}}{y_{q_2}-y_{q_1}} =
\zeta_{q_2 q_1}
 \end{equation}
(Parameters $y_{q}, y_{q_1q_2}$ are listed in Eq. (\ref{parameters}).

One can see, that the Nambu sum rule holds in the form
\begin{eqnarray}
&& [M^{+}_{q_1\bar{q}_2}]^2  + [M^{-}_{q_1\bar{q}_2}]^2 +
[M^{+}_{q_2\bar{q}_1}]^2  + [M^{-}_{q_2\bar{q}_1}]^2\approx 4 [M_{q_1}^2 +
M_{q_2}^2], \nonumber\\&& \quad (q_1\ne q_2); \nonumber\\
&& [M^{P}_{q\bar{q}}]^2  + [M^{S}_{q\bar{q}}]^2 \approx 4 M_{q}^2
\end{eqnarray}

In the case when the t - quark contributes to the formation of the given
scalar excitation,
its mass dominates, and in each channel ($t\bar{t}, t\bar{c}$, ...) we come
to the relation
\begin{equation}
 \sum M_{H, i}^2  \approx 4 M^2_t\,,
  \label{NSR}
\end{equation}
where the sum is over the scalar excitations in the given channel.

The symmetry breaking pattern of the considered model is $U_{L,1}(2)\otimes
U(2)_{L,2} \otimes U(2)_{L, 3} \otimes U(1)_u \otimes ... \otimes U(1)_b
\rightarrow U(1)_u\otimes ...\otimes U(1)_t\otimes U(1)_b$. Among the
mentioned Higgs bosons there are 12 Goldstone bosons that are exactly
massless (in the channels $t(1\pm \gamma^5)\bar b, t \gamma^5\bar{t},
c(1\pm\gamma^5)\bar{s}, c\gamma^5\bar{c}, u(1\pm \gamma^5)\bar{d},
u\gamma^5\bar{u},b\gamma^5\bar{b}, s\gamma^5\bar{s}, d\gamma^5\bar{d}$).
There are Higgs bosons with the masses of the order of the t-quark mass ($
t(1\pm \gamma^5)\bar b, t \bar{t},  t(1\pm\gamma^5)\bar{s},
t\gamma^5\bar{c}, t(1\pm \gamma^5)\bar{d}, t\gamma^5\bar{u}$). The other
Higgs bosons have masses much smaller than the t - quark mass.
A lot of physics is to be added in order to make this model realistic. In
particular, extra light Higgs bosons should be provided with the masses of
the order of $M_t$.

\section{Nambu sum rules in dense quark matter}

In dense quark matter with $\mu > \Lambda_{QCD}$ there may appear several
phases with
different diquark condensates. For example, in the color -
flavor locking phase (CFL) in the framework of the phenomenological model
with
three massless quarks $u,d,s$ the condensate has the form
\cite{CFL0, CFL}
\begin{equation}
\langle [\psi^{i}_{\alpha}]^t i \gamma^2\gamma^0 \gamma^5 \psi^j_{\beta}
\rangle \sim \Phi^I_J \, \epsilon_{\alpha \beta J} \epsilon^{ijI} \sim
(\beta V)^{1/2} C \,  \, \epsilon_{\alpha \beta I} \epsilon^{ijI}
\end{equation}

The symmetry breaking pattern is $SU(3)_L \otimes SU(3)_R
\otimes SU(3)_F\otimes U(1)_A\otimes U(1)_B \rightarrow SU(3)_{CF}$.
There are $36$ scalar and pseudoscalar fluctuations of $\Phi$ around this
condensate
\cite{diquarks}). Among them there are $9+9$ massless Goldstone modes.  The
remaining $9+9$
Higgs modes contain two octets of the traceless modes and two singlet
trace modes. The quark excitations also form singlets and octets. There are
two fermionic gaps (for the octet and for the singlet) $\Delta_{1} = 2
\Delta_{8}$ (Sect. 5.1.2. of \cite{CFL}).
The scalar singlet and octet masses are $M_1 = 2 \Delta_1,
M_8 = 2 \Delta_8$. This may be derived from the results presented in
\cite{m1,m1m8}.

 We already mentioned in the introduction, that in the Hadronic phase the
NJL approximation leads to the Nambu sum rule in the trivial form
$M_{\sigma} = 2 M_{quark}$. However, at nonzero $\mu << M_{quark}$ the Nambu
sum rule in the nontrivial form appears for the diquark states. Namely, the
following values of the masses of the diquarks are presented in  Eq. (46) of
\cite{diquarks}:
\begin{equation}
M_{\Delta}^2 = (2 M_{quark} - 2 \mu)^2;\quad M_{\Delta^*}^2 = (2 M_{quark} +
2 \mu)^2
\end{equation}
So that
\begin{equation}
M^2_{\Delta}+M_{\Delta^*}^2 \approx 2\times 4 M_{quark}^2\, {\rm at}\, \mu
<< M_{quark}
\end{equation}
(Here $\Delta$ is the diquark while $\Delta^*$ is the antidiquark.)

\section{ Veltman identity}

 In the case of the single Higgs boson and in the
absence of the gauge fields the quadratic divergences in the mass
of the Higgs boson vanish if
$3  M_{H}^2  = 4 \sum_f M^2_f$ (see
\cite{Alberghi2008,FrolovFursaev1998,FFZ2003,Zeldovich1968,veltman}).
Here $M_H$ is the scalar boson mass, while the sum
is over the fermions of the model. For the model with triply degenerated
quarks,
this relation is reduced to $
M_{H}^2  = 4 \sum_f M^2_f$.
It looks similar to Eq. (\ref{NSR_}). Nevertheless, their origins are
different. This follows from the fact that the
cancellation of quadratic divergences relies on the identity $N_C=3$ while
the Nambu sum
rule Eq. (\ref{NSR_}) in the models considered above works for any number of
fermion colors. Besides, the number of the components of the scalar
is relevant for Veltman relation. Therefore, its nature  differs from the
nature of the Nambu sum rule.

\section{Conclusions and discussion}

In this paper we consider the bosonic spectrum of various NJL models: from
the condensed matter models of superfluidity to the relativistic models of
top quark condensation. In each case the Nambu sum rule takes place that
relates the masses (or, energy gaps) of the bosonic excitations with the
mass (energy gap) of the heaviest fermion that
contributes to the formation of the given composite scalar boson. (It is
implied that its mass is essentially larger than the masses of the other
fermions
that contribute to the given composite boson.)

We suppose that the top quark contributes to the
formation of the composite Higgs bosons.  There may also appear the other
composite Higgs bosons, whose formation is not related to the top quark.
These Higgs bosons would be light. Since such states are not observed,
 their formation is to be suppressed. Some physics is to be added in order
to provide this. For example, these
 bosons may be eaten by some extra gauge fields that acquire
masses due to the Higgs mechanism.

 The results presented in this paper belong
to the NJL - like models considered in weak coupling. In any realistic
models this is only an approximation.
In QCD the use of the NJL approximation is limited at low energies, in
particular, because confinement is not taken into account. However, the
unknown theory, whose low energy approximation may have the
form of the NJL model, should provide chiral symmetry breaking but cannot be
confining. (Otherwise all quarks would be confined to the regions of space
smaller than TeV$^{-1}$.) This justifies the use of this technique. Besides,
Eq. (\ref{NSR_}) being derived using the NJL approximation does not
contain the parameters of the NJL model: neither the coupling entering the
four - fermion terms nor the cutoff.
As Nambu noticed in \cite{Nambu}, his sum rule may work better than
the NJL approximation itself.

\section*{Acknowledgements}
The authors kindly acknowledge useful remarks by J.D. Bjorken, D.I.
Diakonov,
T.W.B. Kibble, F.R. Klinkhamer, S.Nussinov, M.I. Polikarpov,  A.M. Polyakov,
M.
Shaposhnikov, V.I. Shevchenko, T. Vachaspati, M.I. Vysotsky, V.I. Zakharov.
The
authors are very much obliged to V.B.Gavrilov and M.V.Danilov for the
explanation of the experimental situation with the search of new particles
at
the LHC. This work was partly supported by RFBR grant 11-02-01227, by the
Federal Special-Purpose Programme 'Human Capital' of the Russian Ministry of
Science and Education, by Federal Special-Purpose Programme 07.514.12.4028.
GEV
acknowledges a financial support of the Academy of Finland and its COE
program,
and the EU  FP7 program ($\#$228464 Microkelvin).


\begin{thebibliography}{99}






\bibitem{Nambu}
Yoichiro Nambu,
"Fermion - boson relations in BCS type theories",
Physica D {\bf 15},  147--151 (1985);
"Energy gap, mass gap, and spontaneous symmetry breaking",
in: {\it BCS: 50 Years}, eds. L.N. Cooper and D. Feldman, World Scientific
(2010).



\bibitem{NJL}
Y. Nambu, G. Jona-Lasinio,
"Dynamical model of elementary particles based on an analogy with
superconductivity. I,"
Phys. Rev. {\bf 122}, 345--358 (1961).

\bibitem{Englert}
F. Englert, R. Brout,
"Broken Symmetry and the Mass of Gauge Vector Mesons",
Phys. Rev. Lett.  {\bf 13}, 321--23    (1964).

\bibitem{Higgs}
P. Higgs,
"Broken Symmetries and the Masses of Gauge Bosons",
Phys. Rev. Lett. {\bf 13}, 508--509 (1964).

\bibitem{Kibble}
G. Guralnik, C.R. Hagen, T.W.B. Kibble,
 "Global Conservation Laws and Massless Particles",
Phys. Rev. Lett. {\bf 13}, 585--587 (1964).

\bibitem{CMSHiggs}
"Search for the standard model Higgs boson produced in association with W
and Z bosons in pp collisions at s‰Ãû=7 TeV",  CMS Collaborati
n
arXiv:1209.3937 ; CMS-HIG-12-010 ; CERN-PH-EP-2012-253.

"Observation of a new boson at a mass of 125 GeV with the CMS experiment at
the LHC", CMS Collaboration, arXiv:1207.7235; CMS-HIG-12-028;
CERN-PH-EP-2012-220.- Geneva : CERN, 2012 - 59 p. - Published in : Phys.
Lett. B 716 (2012) 30-61

"A search for a doubly-charged Higgs boson in pp collisions at s‰Ãû = 7 TeV
  CMS Collaboration
arXiv:1207.2666 ; CMS-HIG-12-005 ; CERN-PH-EP-2012-169. - 2012. - 39 p.



\bibitem{ATLASHiggs}
"Observation of a new particle in the search for the Standard Model Higgs
boson with the ATLAS detector at the LHC",  ATLAS Collaboration,
Phys.Lett. B716 (2012) 1-29, CERN-PH-EP-2012-218,
arXiv:1207.7214 [hep-ex]


\bibitem{CMSexc}
"Observation of a new boson with a mass near 125 GeV", The CMS
Collaboration,
CMS PAS HIG-12-020, available at CERN information server as
http://cdsweb.cern.ch/record/1460438/files/HIG-12-020-pas.pdf



\bibitem{ATLASexc}
http://cdsweb.cern.ch/record/1470512/files/ATL-PHYS-SLIDE-2012-459.pdf







\bibitem{CDF} CDF Collaboration, CDF/PUB/EXOTICS/PUBLIC/10603, July 2011

\bibitem{CMS} CMS Collaboration, Phys. Rev. Lett. 108, 111804 (2012),
arXiv:1202.1997

\bibitem{325}
Krzysztof A. Meissner, Hermann Nicolai,
"A 325 GeV scalar resonance seen at CDF?",
arXiv:1208.5653.

\bibitem{320}
L. Maiani, A. D. Polosa, V. Riquer,
"Probing minimal supersymmetry at the LHC with the Higgs boson masses",
 arXiv:1202.5998.









\bibitem{ATLAS}
ATLAS Collaboration,
Phys. Lett. B {\bf 710}, 49--66 (2012).

 http://atlas.ch/news/2011/Higgs-note.pdf


\bibitem{VolovikZubkov2012}
G.E. Volovik and M.A. Zubkov,
"The Nambu sum rule and the relation between the masses of composite Higgs
bosons",
  arXiv:1209.0204.

\bibitem{Miransky}
V.A. Miransky,  Masaharu Tanabashi, Koichi Yamawaki,
``Dynamical electroweak symmetry breaking with large anomalous
dimension and t quark condensate",
Phys. Lett. B {\bf 221}, 177--183 (1989);
``Is the t quark responsible for the mass of W and Z bosons?",
Mod.  Phys. Lett. A {\bf 4}, 1043--1053  (1989).

\bibitem{topcolor1}
William A. Bardeen, Christopher T. Hill, Manfred Lindner,
"Minimal Dynamical Symmetry Breaking of the Standard Model,"
Phys. Rev. D {\bf 41}, 1647--1660 (1990).

\bibitem{VollhardtWolfle1990}
D. Vollhardt  and P.  W\"olfle,
{\it The superfluid phases of helium 3},  Taylor and Francis, London
(1990).

\bibitem{Gazit2012}
Snir Gazit, Daniel Podolsky, Assa Auerbach,
Fate of the Higgs mode near quantum criticality,
arXiv:1212.3759

\bibitem{Barlas2012}
Yafis Barlas, C. M. Varma,
Higgs Bosons in D-wave Superconductors,
arXiv:1206.0400.

\bibitem{KunChen2013}
Kun Chen, Longxiang Liu, Youjin Deng, Lode Pollet, Nikolay Prokof'ev,
Universal properties of the Higgs resonance in (2+1)-dimensional U(1) critical systems,
arXiv:1301.3139.

\bibitem{Lee1988}
R. Movshovich, E. Varoquaux, N. Kim, and D.M. Lee,
"Splitting of the squashing collective mode of superfluid $^3$He-B by a
magnetic field",
Phys. Rev. Lett. {\bf 61}, 1732--1735 (1988).

\bibitem{Collett2012}
C. A. Collett, J. Pollanen, J. I. A. Li, W. J. Gannon, W. P. Halperin,
"Zeeman splitting and nonlinear field-dependence in superfluid $^3$He",
arXiv:1208.2650.

\bibitem{VolovikVachaspati1996}
G.E. Volovik and T. Vachaspati,
"Aspects of $^3$He and the  standard electroweak model,"
Int. J. Mod. Phys. {\bf B~10}, 471--521 (1996);
cond-mat/9510065.

\bibitem{BrusovPopov1980}
P.N. Brusov and V.N. Popov,
"Zero-phonon branches of the Bose spectrum in the A phase of a system of the
He3 type",
JETP {\bf 52}, 945--949 (1980).

\bibitem{VolovikKhazan1982}
G.E. Volovik, M.V. Khazan,
 "Dynamics of the A-phase of  $^3$He at low pressure,"
JETP {\bf 55},  867--871 (1982);
"Classification of the collective modes of the order parameter in superfluid
$^3$He," JETP {\bf 58}, 551--555  (1983).


\bibitem{BrusovPopov1981}
P.N. Brusov and  V.N. Popov,
"Superfluidity and Bose excitations in He3 films",
JETP {\bf 53}, 804--810 (1981).

\bibitem{2HMC}
Cristian Valenzuela Roubillard, PhD thesis, arXiv:hep-ph/0503289;
Cristian Valenzuela, Phys. Rev. D {\bf 71}, 095014 (2005);
arXiv:hep-ph/0503111.

\bibitem{2HMMiransky}
Michio Hashimoto, V.A. Miransky,
"Dynamical electroweak symmetry breaking with superheavy quarks and 2+1
composite Higgs model",
  Phys. Rev. D {\bf 81}, 055014 (2010); arXiv:0912.4453.





\bibitem{Z2013}
M.A.Zubkov,  "Gauge theory of Lorentz group as a source of the dynamical
electroweak symmetry breaking", arXiv:1301.6971

\bibitem{diquarks}
D. Ebert, K.G. Klimenko, V.L. Yudichev,
Eur. Phys. J. C {\bf 53}, 65--76 (2008).

\bibitem{CFL0}
Mark Alford, Krishna Rajagopal, Frank Wilczek,
Nuclear Physics B {\bf 537}, 443--458 (1999).

\bibitem{CFL}
Michael Buballa, Phys.Rept. {\bf 407},  205--376 (2005).

\bibitem{m1}
Roberto Anglani, Massimo Mannarelli, Marco Ruggieri,
"Collective modes in the color flavor-locked phase",
New J. Phys.{\bf 13}, 055002 (2011).

\bibitem{m1m8}
Minoru Eto, Muneto Nitta, and Naoki Yamamoto,
Phys. Rev. Lett.  {\bf 104}, 161601 (2010);
Minoru Eto and Muneto Nitta,
Phys. Rev. D {\bf 80}, 125007 (2009).



\bibitem{Alberghi2008}
G. L. Alberghi, A. Y. Kamenshchik, A. Tronconi, G. P. Vacca, G. Venturi,
"Vacuum energy, cosmological constant and Standard Model physics",
Pis'ma  ZhETF {\bf 88}, 819--824 (2008);
arXiv:0804.4782.


\bibitem{FrolovFursaev1998}
V.P. Frolov and D. Fursaev,
"Black hole entropy in induced gravity: Reduction to 2D quantum field theory
on the horizon",
Phys. Rev.  {\bf D~ 58}, 124009  (1998)
[arXiv:hep-th/9806078].

\bibitem{FFZ2003}
V. Frolov, D. Fursaev and A. Zelnikov,
"CFT and black hole entropy in induced gravity",
 JHEP {\bf 0303} (2003) 038
 [arXiv:hep-th/0302207].

\bibitem{Zeldovich1968}
Ya. B. Zel'dovich,
"The cosmological constant and the theory of elementary particles",
Usp. Fiz. Nauk {\bf 95}, 209--230 (1968)
[Sov. Phys.-Uspekhi {\bf 11}, 381--393 (1968)].

\bibitem{veltman}
M. Veltman,
``The infrared-ultraviolet connection,''
Acta Physica Polonica B {\bf 12}, 437--457 (1981).


\end{thebibliography}
\end{document}